# Strategy in a Digital World

**Melinda D'Cruz**
Information Systems School
Queensland University of Technology
Brisbane, Australia
Email: melinda.dcruz@student.qut.edu.au

**Greg Timbrell**
Information Systems School
Queensland University of Technology
Brisbane, Australia
Email: g.timbrell@qut.edu.au

**Jason Watson**
Information Systems School
Queensland University of Technology
Brisbane, Australia
Email: ja.watson@qut.edu.au

## Abstract

Organizations are increasingly adopting digital strategies and investing heavily in digital technologies and initiatives. However, to date, there does not appear to be a clear understanding of digital strategies and their purpose, which forms the motivation for this research. This research-in-progress study aims to address this research gap by exploring 1) the various conceptions of digital strategy, and 2) the way in which digital strategies differ from conventional strategies. We interviewed three senior executives and employed thematic analysis to analyse the interview data, which resulted in the construction of ten themes that were grouped under three theoretical constructs. We then explored the applicability of the six dimensions of strategy proposed by Hax (1990) in the digital context and proposed two additional dimensions. The contribution of this study is to provide a deeper understanding of digital strategy to support further academic research and provide guidance to practitioners.

# 1 INTRODUCTION

'Digital' appears to be a phenomenon of growing importance at the board level as evidenced by the birth of a new role in the C-Suite, a Chief Digital Officer (CDO). In fact, 52% of Chief Executive Officers and other senior executives surveyed by Gartner in 2013 claimed that their organizations have in place a 'digital strategy' (McGee 2013). Organizations are investing heavily in digital technologies and initiatives, with Gartner (2014) predicting an increase in technology spending of 7.4% in the Asia Pacific region in 2015 totalling US$811 billion in an effort to "embrace the digital economy". There are currently several academic studies on digital strategy: Bharadwaj et al. (2013) explore the scope, scale and speed of digital business strategies and sources of value creation and capture; Mithas et al. (2013) examine the influence of the industry environment and digital strategy posture on digital business strategy; Grover and Kohli (2013) discuss the desirability and caveats in embracing digital business strategies; and, Pagani (2013, p. 617) investigates "the dynamic cycle of value creation and value capture points in digitally enabled networks". Yet, despite significant organizational investments in digital technologies and various academic studies, to date, a unified understanding of the phenomenon of digital strategy does not exist, which is the motivation for this research-in-progress paper. The contribution of this paper will be a deeper understanding of digital strategies, which will provide guidance in academic research and facilitate more effective strategy derivation in industry.

This research appears to be the first academic study to explore qualitatively the different ways in which entities conceive digital strategies. A common and unified understanding of digital strategies and the way in which they differ from conventional strategies is paramount and a necessary precursor to developing prescriptive frameworks for formulating and implementing digital strategies. By examining existing literature and drawing on three interviews with senior executives in government, and large, global professional service firms, we have begun to address this research gap by investigating the following research questions: *1) What are the various conceptions of digital strategy? 2) How does digital strategy differ from conventional strategy?* While the broader research study aims to explore the way in which organisations define, develop and implement digital strategies, the purpose of this paper is not to examine specific approaches to digital strategy development and implementation or propose frameworks for effective strategy derivation. Additionally, while business models and strategy are related constructs, this paper does not aim to explore the concept of business models in detail or examine the research findings in the context of emerging business models.

In this paper, we firstly present the key concepts and findings from the preliminary literature review, followed by the theoretical lens and research methods adopted in this study. We then discuss the initial findings in relation to the conceptions of digital strategy based on the interview data. We apply this understanding of digital strategy and evidence from existing literature to the six dimensions of strategy identified by Hax (1990) in order to explore the way in which digital strategies may differ from conventional strategies. Finally, we conclude by summarizing the study's contributions and limitations, and opportunities for further research.

# 2 LITERATURE REVIEW

## 2.1 Strategy in Business Literature

The existing body of literature on strategy is highly contentious and as Whittington (1993) highlights, there is limited consensus on the definition of and approaches to strategy. Chandler (1962, p. 13) defined strategy as "the determination of the basic long-term goals and objectives of an enterprise, and the adoption of courses of action and the allocation of resources necessary for carrying out these goals". However, according to Mintzberg (1978, p. 935), such notions of strategy imply that the phenomenon is "(a) explicit, (b) developed consciously and purposefully, and (c) made in advance of the specific decisions to which it applies." Referring to this as "intended strategy", he argues that strategies may also emerge as a response to environmental changes, known as "emergent strategy" (Mintzberg 1978, p. 935; 1987, p. 68).

Hax (1990) proposed a comprehensive definition that unified differing notions of strategy, claiming that the phenomenon comprises six main dimensions:

1. Strategy is "a means of establishing an organization's purpose in terms of its long-term objectives, action programs, and resource allocation priorities" (Hax 1990, p. 35)
2. Strategy is "a coherent, unifying, and integrative pattern of decisions" (Hax 1990, p. 34)

3. Strategy is "a definition of a firm's competitive domain" (Hax 1990, p. 35)

4. Strategy is "a response to external opportunities and threats and to internal strengths and weaknesses as a means of achieving [long-term, sustainable] competitive advantage" (Hax 1990, p. 35)

5. "Strategy is a definition of the economic and non-economic contribution the firm intends to make to its stakeholders" (Hax 1990, p. 36)

6. Strategy is "a logical system for differentiating managerial tasks at corporate, business, and functional levels" (Hax 1990, p. 36)

The next section explores the notion of strategy from an Information Technology (IT) perspective.

## 2.2　Strategy in Information Technology Literature

According to Weill and Broadbent (1998, p.24), an entity's IT portfolio represents "… its entire investment in information technology, including all the people dedicated to providing information technology services, whether centralized, decentralized, distributed, or outsourced. The investments include all computers, telecommunications networks, data, software, training, programmers, support personnel, point-of-sale systems, databases…".

Traditionally, IT strategy was perceived as a functional strategy that supported the business through cost savings and improved efficiency (Burg and Singleton 2005). However, Henderson and Venkatraman (1993, p. 472) assert that IT "is transcending its traditional 'back office' role and is evolving toward a 'strategic' role with the potential not only to support chosen business strategies, but also to shape new business strategies". Perhaps the emergence of e-business and e-commerce facilitated by the first generation of the web ("Web 1.0"), characterised predominately by static websites providing limited interactivity (Aghaei et al. 2012), may be an example of IT shaping business strategy.

There is little dispute that IT and business strategies should be aligned in order for IT to provide value to the business, increase sales and profit, provide competitive advantage, offer flexibility to respond to new opportunities that arise and facilitate an effective business model (Avison et al. 2004; Henderson and Venkatraman 1993; Tallon and Pinsonneault 2011; Tan and Gallupe 2006; Venkatraman et al. 1993). The landmark Strategic Alignment Model (SAM) provides a way to transform the business to exploit IT-enabled opportunities (Scott Morton 1991) by ensuring functional integration between the business and IT domains and strategic fit between the internal and external dimensions (Avison et al. 2004; Henderson and Venkatraman 1993; Venkatraman et al. 1993). While the fundamental concept of alignment between business objectives and internal IT capabilities pioneered by the SAM remains relevant, it fails to acknowledge the pervasiveness of digital technology in modern times, and the role of digital strategies in enabling business transformation, by exploiting pervasive digital connections and digital assets external to the organisation e.g. infrastructure, platform or software 'as a service'.

The next section explores the phenomenon of digital strategy as presented in the literature.

## 2.3　Digital Innovations, Business Models and Digital Strategy

O'Reilly (2007, pp. 1-16) defines the second generation of the web ("Web 2.0") as "a set of principles and practices", which involve: reaching out "to the entire web" and "leveraging the long tail through customer self-service"; acquiring rich and unique data that is difficult for competitors to harness; innovating by integrating services; encouraging user participation and harnessing intelligence such as through blogs and wikis; providing software as services with continual updates and new functionality introduced regularly; supporting lightweight development models "that allow for loosely coupled systems"; ensuring that software supports various devices and platforms; and providing "rich user experiences".

Recent digital innovations, supported by these Web 2.0 principles and practices, are transforming businesses and social relationships (Bharadwaj et al. 2013, p. 472). In a 2012 McKinsey global survey, Chief Executives Officers identified the top three current trends of strategic importance in 'digital businesses' as "big data and analytics, digital marketing and social-media tools, and the use of new delivery platforms such as cloud computing and mobility" (Brown and Sikes 2012). New digital innovations increasingly threaten the existence of traditional business models and have given rise to new business models. Osterwalder et al. (2005, p. 3) define a business model as "a conceptual tool containing a set of objects, concepts and their relationships with the objective to express the business logic of a specific firm…". Anderson (2009) identifies four new business models built on the concept of 'free' things: direct cross-subsidies (whereby a free product necessitates a purchase), three-party market (whereby there's a free exchange between two entities and a third entity pays to participate in this market), freemium (whereby a paid, premium product is available and is superior to the free version) and nonmonetary markets (whereby items are given away for free). According to Teece (2010, p. 180), a "business model is more generic than a business strategy" and strategy analysis is a crucial element "in designing a competitively sustainable business model".

A digital business strategy is thought to represent a "fusion" between the business and IT strategies and is claimed to be an "organizational strategy formulated and executed by leveraging digital resources to create differential value" (Bharadwaj et al. 2013, p. 472). It is argued that while an IT strategy may be classed as a functional-level strategy, a digital business strategy should be viewed as a business strategy in the digital world given organisations' growing dependence on information, and digital connections and communications (Bharadwaj et al. 2013). However, the difference between an IT strategy and a digital strategy is unclear in existing literature. Further research will be undertaken as part of the current study to explore the difference between an IT strategy and a digital strategy.

## 3  RESEARCH METHOD

### 3.1  Philosophical and Theoretical Framework

The sub-sections below outline the philosophical and theoretical assumptions underpinning this study.

#### 3.1.1  Purpose of the study

A research study may be considered exploratory, descriptive or explanatory (Neuman 2006). This study is predominately exploratory given it seeks to explore the various conceptions that entities possess of digital strategy and the limited academic research that currently exists on the phenomenon.

#### 3.1.2  Ontology

As Blaikie (1993, p. 6) and Crotty (2003, p. 10) acknowledge, ontology is the "study of being". Ontological positions range from realism to idealism (Blaikie 2007; Ormston et al. 2014) with the former entailing a belief in an 'external reality' that cannot be influenced by human activity whilst the latter involves the belief that reality does not exist aside from human understanding and interpretation (Blaikie 2007; Ormston et al. 2014). This study adopts the ontological position of realism as it may be argued that technology exists independent of human thought and consciousness. For instance, customers may purchase books through Amazon without considering the gamut of technologies that facilitate this activity including the hardware (e.g. cables, routers and servers), software, databases, and protocols (e.g. TCP/IP). It is not to say that these technologies do not exist; rather it may be argued that different people experience and perceive a phenomenon differently and possess different levels of awareness (Edwards 2007).

Similarly, while a specific instance or execution of a digital strategy may be influenced by entities' thoughts and ideas, it may be argued that the range of possible strategic choices or courses of action that are available to execute digital strategies will exist regardless of human consciousness and thought. Furthermore, as strategy represents a "pattern of decisions" (Hax 1990, p. 34) that may be intended or emergent, it may be argued that strategy exists regardless of whether entities deliberately consider and employ a certain strategy. Therefore, the study assumes that digital strategy is an external reality that will exist regardless of human thought and activity. According to Crotty (2003, p. 10), ontology sits "… alongside epistemology informing the theoretical perspective, for each theoretical perspective embodies a certain way of understanding *what is* (ontology) as well as a certain way of understanding *what it means to know* (epistemology)".

#### 3.1.3  Epistemology

There are three main epistemological positions: Objectivism, constructionism, and subjectivism (Crotty 2003). In the objectivist view, objects possess 'intrinsic meaning' and it is possible to "discover the objective truth" whereas in the subjectivist view, objects do not possess 'intrinsic meaning' and the subject imposes meaning on the object by means other than interacting with the object (Blaikie 2007, pp. 18-19). In contrast, constructionism maintains that "meanings are constructed by human beings as they engage with the world they are interpreting" (Crotty 2003, p. 43). This study assumes an epistemological position of social constructionism, which refers to "the collective generation and transmission of meaning" (Crotty 2003, p.58). As Crotty (2003, p. 43) accentuates, a 'Tree' "is likely to bear quite different connotations in a logging town, an artists' settlement and a treeless slum". Similarly, digital strategy may have different connotations depending on the social context in which it is interpreted and understood. While the phenomenon of digital strategy may be considered an external reality (ontology), it can be investigated by examining the way in which entities construct meaning by collectively executing digital strategies in specific social and organisational contexts (epistemology). The theoretical perspective or research paradigm adopted in this study embodies this view of social reality as discussed in the next section.

### 3.1.4 Research Paradigm

Three main approaches to social science are identified: namely, the positivist approach, the interpretive approach and the critical approach (Crotty 2003; Neuman 2006). The current study most closely aligns with the interpretive paradigm as it aims to identify the various conceptions of digital strategy and assumes an inductive approach to reasoning, adopting the view that perceptions of digital strategy are constructed by entities experiencing the phenomenon in specific social contexts.

### 3.1.5 Research Use

Neuman (2012) distinguishes between two main types of research each with differing uses: Basic Research and Applied Research. The current study is predominately basic research as it primarily seeks to contribute to the existing theoretical body of knowledge on digital strategies and strategy in general, with a strong emphasis on scientific and methodological rigour. Additionally, this paper provides a significant contribution to industry as a clear and comprehensive understanding of digital strategy and the way in which it differs from conventional strategy will prove invaluable to consulting firms and other organisations that seek to develop and implement digital strategies.

### 3.1.6 Approach to Data Collection and Analysis

Given that the current study is primarily exploratory with a focus on generating theory, and aims to investigate the diverse conceptions of digital strategy, it would appear that qualitative methods of data collection and analysis assuming an inductive and ideographic approach are the most appropriate.

## 3.2 Data Collection

By applying the four points of reference proposed by Flick (2006), semi-structured interview was adopted as the data collection method for this exploratory study as it facilitates the collection of rich, abundant and detailed empirical evidence, and provides the versatility to explore emerging concepts as well as further investigate findings from existing literature.

The interviews were structured as a three-part process. The first set of questions aimed to gather information that may be used to contextualize participant responses. The second set of questions explored participants' understanding of digital strategies whereby the interviewer(s) did not attempt to influence the respondent by introducing concepts from existing literature as this may defeat the exploratory purpose of posing this set of questions by stifling the interviewee from providing novel insights into the phenomenon. The final set of questions aimed to elicit participants' conceptions on aspects of digital strategies as presented in literature. During this process, 'active listening' strategies were adopted, whereby the investigator sought to identify the underlying or implicit meaning in the interviewee's response through probing questions (Liamputtong 2013; McCracken 1988). In this regard, meaning was constructed through the interaction between the interviewer and interviewee.

Three participants were interviewed regarding their perceptions of digital strategies. Participant A is a Managing Director of the Australian Digital Practice within a global consulting firm. This participant has extensive prior experience in various roles including Head of Mobility, Consultant and Academic. Participant B is a Business Solution Executive within the Business Processing Practice in a global technology firm that also provides consulting services. This participant is currently involved in the initial strategy development phases with a background in implementing digital programs of work predominately within financial institutions. Participant C is a senior government executive with extensive experience in delivering large technology solutions ranging from core backend processing systems to front-end digital solutions.

## 3.3 Data Analysis

The data analysis process employed in this study consists of two main phases: Preparation, and Thematic Analysis. The preparation phase involved transcribing the audio-recorded interviews, becoming familiar with the data and establishing procedures for the secure storage and analysis of the data. As thematic analysis is a "foundational method for qualitative analysis" that is highly compatible with semi-structured interviews, it was employed in this study (Braun and Clarke 2006, pp. 78-79).

Whittington (1993) conceives strategy in relation to two main dimensions: (1) the outcomes produced by strategy and (2) the way in which it is implemented. We extend this concept and propose that digital strategy may be understood in relation to three main dimensions or theoretical constructs: (1) the purpose of and outcomes produced by digital strategy, (2) the process of executing digital strategies, and (3) the relationship between digital strategies and existing conventional strategies.

A cutting and sorting technique was used to analyse the interview data (Ryan and Bernard 2003). Data relevant to the research questions were identified and sorted into similar quotes using word processing software. Themes were constructed based on recurring ideas, participants' terms / in Vivo codes, and similarities and differences in participants' expressions (Saldana 2009; Ryan and Bernard 2003), and categorised under the three theoretical constructs.

### 3.4 Research Quality

Recommendations outlined by Guba (as cited in Shenton 2004) were employed in the current study to enhance research rigour. Venkatesh et al. (2013) have categorised various types of validity applicable to qualitative research into one of three groups: design validity, analytical validity and inferential validity. Design validity in this study is demonstrated through: (1) descriptive validity (by including verbatim quotations from the interview data to ensure accurate reporting), (2) transferability (through purposive sampling to investigate the phenomenon in different contexts), and (3) credibility (by exercising reflexivity, adopting well-established data collection and analysis methods, and ensuring investigator triangulation as the Principal Researcher and Research Supervisor facilitated interviews). As this study adopts a constructionist perspective, to be considered credible, it was ensured that the findings adequately reflect the multiple constructed versions of reality provided by the study's participants (Liamputtong 2013). Analytical validity in this study is demonstrated through theoretical validity and plausibility by providing an audit trail, which identifies the quotations from the interview data that supports each identified theme. Inferential validity in this study is demonstrated through interpretive validity and confirmability by providing an audit trail, ensuring investigator triangulation, and exercising researcher reflexivity to minimise bias.

## 4 INITIAL FINDINGS AND ANALYSIS

### 4.1 Conceptions of Digital Strategy

Numerous conceptions of digital strategy were identified from the interview data by employing thematic analysis. The initial findings are summarised below with a subset of supporting quotes from the interviews.

#### 4.1.1 Construct 1: The purpose of / outcomes from digital strategies

**Theme 1: Digital strategies focus on operational process efficiency, enhancing the customer experience and/ or business model transformation**

The interview findings reveal that digital strategies focus on improving operations (e.g. achieving cost savings, improving productivity), enhancing the customer experience (e.g. through real-time communications via social media) and/or achieving business model transformation (e.g. through strategic partnerships, offering new services). Supporting evidence from an interview:

*Interviewer: "What is your understanding of a digital strategy and its purpose?"*

*Participant B: "Ok so I tend to look at digitization in three key pillars of where to focus in regards to digitization strategy. Typically customers will look at firstly, operations so when enabling a digitization strategy they are looking for operational process efficiency which will be cost reductions, reducing their time to market of products etc. ... The second is around customer experience so digitization strategies now is all about the new generation of you know, social media and you know instantaneous... So operations one, customer experience two, the third is around business model. ..."*

**Theme 2: Digital strategies range from digital point solutions to more holistic strategies**

Digital strategies may range from holistic solutions (e.g. which facilitate new business models or business transformation) to point solutions (e.g. a strategy that focuses on mobility). Supporting evidence from an interview:

*Participant A: "Yep... I've got two roles. The first role is an internal facing one... The other one is helping deliver, execute digital strategies... and that varies from... point digital strategies – so what does mobility mean, what to do around mobility, to more holistic strategies that apply... what does*

*this mean for us as a corporation and what does digital mean for us as a business and what we should do about it."*

**Theme 3: Digital strategy relates to the provision of online services**

The interview data indicates that a digital strategy is conceived as relating to the provision of online services to expose information to customers and allow them to update that information thereby facilitating customer self-service:

*Participant C: "I think our IT strategy covers people, process and technology... To me the digital strategy covers aspects of those again, with my definition of digital being around online services."*

**Theme 4: Digital strategy allows entities to sustain their competitive position**

The interview data reveals that entities in many industries are quick to imitate competitors thereby making it difficult for organisations to sustain competitive advantage by achieving differentiation. Therefore, it appears that the purpose of digital strategies for these organisations is to sustain their competitive position as the evidence shows:

*Participant A: "... we found that banks, insurance companies... the digital intensity was high but the level of differentiation of intensity was very low. So everyone was doing the same... thing. In terms of competitive advantage and sustainable competitive advantage, the reality is that the payoff from digital is very low for a lot of people"*

*Interviewer: "So it's not necessarily a competitive advantage but sustaining your competitive position."*

*Participant A: "Correct yes. ..."*

#### 4.1.2 Construct 2: The process of executing digital strategies

**Theme 5: Digital strategy is a response to technological megatrends and consequent social and behavioural changes**

Digital strategy is perceived as the way in which entities respond to two main changes: 1) changes in technology and 2) fundamental changes to the way in which people behave, work, transact, and use and share information. Supporting evidence from an interview:

*Interviewer: "Is this digital strategy so nebulous that I can't even put my hands around it? ..."*

*Participant A: "... all digital really means is the collation of a number of megatrends going on and those megatrends influence industries and companies in their own specific way..."*

*Interviewer: "Let's get back to those megatrends. What are those megatrends that you're..."*

*Participant A: "Technology. ... social, mobile, analytics, cloud... augmentation, robotics... those kind of leading edge technologies that are affecting the way that people consume media and interact with each other. The way people are behaving... people are interacting with technology in a fundamentally different way..."*

**Theme 6: Digital strategy development and execution as an iterative and experimental process**

The interview findings suggest that digital strategy development and execution is a highly iterative and experimental process:

*Participant A: "...I think that the essential nature of digital strategy in one where the technologies themselves make it more amenable to testing and learning. So through the strategy development process, there's a stronger element and emphasis on actually building and creating things, concepts and ideas, testing them with customers, and iterating strategy in a way that doesn't normally take place in corporate strategy development..."*

**Theme 7: Digital strategy development and execution should be agile and responsive**

The interview data reveals that a significant aspect of developing and executing a digital strategy is the ability of an organisation to maintain agility and be responsive to customer needs:

*Participant C: "... the essence of a digital strategy is the service they provide and the speed in which you can provide those services, so speed to market around your online services, your digital services is paramount. ... that expectation of immediacy... you've got to have the agility and the flexibility and the responsiveness to be able to meet that ever-changing and ever-growing demand..."*

#### 4.1.3   Construct 3: The relationship between digital and conventional strategies

**Theme 8: Digital strategy is an enabler of corporate, business and/or functional strategies**

The interview data reveals that a digital strategy may be an enabler of the corporate, business and/or functional strategies:

*Participant A: "… digital isn't a corporate strategy digital… [will be an] enabler of or an accelerant or a creator of other options within corporate strategy."*

*Participant B: "Digital to me means a new style of business. Digital is a new style and is an enabler for businesses of the future."*

**Theme 9: Digital strategy is displacing the IT strategy**

The interview data reveals that increasingly IT strategies involve supporting legacy systems and maintaining the status-quo while transformational initiatives are undertaken as part of digital strategies:

*Interviewer: "… are you finding that digital strategy is, you know that IT strategy is disappearing from the vocabulary in organisations and digital strategy is replacing it?"*

*Participant B: "Yeah, I think so. I think IT now is becoming, the view of IT is becoming I guess traditionally under a CIO, you know, keeping the lights on, the legacy, the old world. Digital tends to be the new world. So yeah it does seem to be moving toward that way and typically IT budgets are focused purely on maintaining the status quo and the digital investments seem to be all about transforming."*

**Theme 10: Digital strategy is a subset of the IT strategy**

Contrary to the previous theme, the interview data also indicates that a digital strategy is a component of a broader IT strategy.

*Participant C: "… So we have an overarching business strategy – what do we want to be as a business? IT strategy says this is how we can help you be what you want… Digital would be part of that IT strategy."*

#### 4.1.4   Summary

Further interviews will be conducted to elicit additional conceptions of digital strategy to the point of theoretical saturation. While Theme 9 appears to suggest that digital strategy is partly replacing the IT strategy, Theme 10 indicates that a digital strategy is a component of a broader IT strategy, which appears to be contradictory. Additional research is necessary to investigate the social contexts in which these conceptions emerged. Perhaps this may be attributable to entities in government organisations possessing a fundamentally different view of digital strategy as opposed to entities in professional service firms.  The positioning of a digital strategy in relation to conventional strategies is also nebulous as digital strategy appears to be an enabler for all levels of strategy in an organisation, indicating that it may be embedded into various strategies. This may present new governance challenges, providing opportunities for further academic research.

Existing literature was examined to further investigate the conceptions of digital strategy outlined above. In the next section, we explore the conceptions of digital strategy based on the interview data and existing literature through the lens of the six dimensions of strategy proposed by Hax (1990) in order to explore the way in which digital strategies differ from conventional strategies.

### 4.2   Discussion using Hax's Dimensions of Strategy

#### 4.2.1   Strategy as a Means of Establishing Purpose

Hax (1990, p. 35) asserts that strategy is "a means of establishing an organization's purpose in terms of its long-term objectives, action programs, and resource-allocation priorities". We translate this dimension of strategy to the context of digital strategies as follows: **Digital strategy is a means of establishing objectives, action programs and resource-allocation priorities for digitization**. This may involve setting high-level strategic objectives and measures for goals such as improved operational efficiency, enhanced customer experience and business model transformation (Participant B), which aligns with findings from a study conducted by MIT Center for Digital Business and Capgemini Consulting (2011).

#### 4.2.2 Strategy as an Integrative Pattern of Decisions

Hax (1990, p. 34) states that strategy is a "coherent, unifying, and integrative pattern of decisions". It may be argued that digital strategy is also a 'coherent, unifying, and integrative pattern of decisions' that may be planned in advance prior to implementation (Refer to the previous dimension of strategy), or emerge as a response to fundamental technological, social and behavioural changes (Refer to Theme 5 in the previous section).

#### 4.2.3 Strategy as a Definition of a Firm's Competitive Domain

Hax (1990, p. 35) claims that strategy involves "defining the businesses a firm is in or intends to be in". However, digital platforms allow "firms to break traditional industry boundaries and to operate in new spaces and niches that were earlier only defined through those digital resources" (Bharadwaj et al. 2013, p. 474). Furthermore, according to Participant A, uncoupling "*information and technology assets*" and exposing assets through application programming interfaces provide tremendous opportunities in the supply chain. Furthermore, as Bharadwaj et al. (2013, p. 474) purport, digital business strategies extend beyond supply chains and traditional firm and industry boundaries "to loosely coupled dynamic ecosystems" comprising "the business ecosystem, alliances, partnerships, and competitors". This concept may be extended to include customers, because as Participant A highlights, the business model for companies such as Facebook seek to monetize customer information. Thus, we propose the following as a more representative dimension of digital strategies in organisations: ***Digital strategy defines the dynamic ecosystem in which an entity operates or intends to operate***.

#### 4.2.4 Strategy as a Means of Achieving Sustainable Competitive Advantage

According to Hax (1990, p. 35), strategy is "a response to external opportunities and threats and to internal strengths and weaknesses" to achieve "long-term sustainable advantage". As McQuivey (2013) highlights, digital technologies and platforms reduce barriers to market entry particularly as they facilitate new business models built on 'free things', which render cost leadership strategies less relevant. In addition, characteristics of Web 2.0 such as "innovation in assembly" allow novel products to be created such as through mash-ups and integration of available services thereby increasing the threat of substitutes (O'Reilly 2007, p. 13). As Grover and Kohli (2013, p. 660) purport, "Today's competitive advantage is based on a succession of short-term advantages through digital initiatives that are a part of broader…" digital business strategies. Therefore, we propose the following as a more representative dimension of digital strategies: ***Digital strategy is a response to external opportunities and threats and to internal strengths and weaknesses to achieve competitive advantage through successive ephemeral advantages.***

However, digital may also be perceived as a way for entities to sustain their competitive position (Refer to Theme 4 in the previous section). This leads us to consider an alternative view of this dimension, which requires additional investigation to be verified: ***Digital strategy is a response to external opportunities and threats and to internal strengths and weaknesses in order to sustain an entity's competitive position.***

#### 4.2.5 Strategy as a Definition of a Firm's Planned Contribution to Stakeholders

Hax (1990, p. 36) indicates that strategy "is a definition of the economic and non-economic contribution the firm intends to make to its stakeholders". Customers appear to be a key stakeholder in an entity's decision to adopt digital strategies as entities seek to enhance the customer experience by better understanding and engaging with their customers (Refer to Theme 1 in the previous section). Consumers also emerge as co-creators of value (e.g. by contributing digital artefacts such as photos on Instagram) as digital public goods are based on the 'Prosumer model' (Rosemann et al. 2011). Digital public goods, according to Rosemann et al. (2011) are easy to access, often free for use, intuitive to consume, and offer greater benefits to users as the community grows and consumption increases. Additionally, business models such as Airbnb and Uber are facilitated by crowdsourcing, and platforms such as Facebook rely on crowdsourcing of application development (O'Reilly 2007).

Employees and entities in the supply chain also emerge as significant stakeholders. According to Participant A, "*it's more of… using digital not just for customers but for employees, the whole supply chain doing things differently and we call it service design…*". Therefore, it may be surmised that the specific contribution of digital strategies to stakeholders will depend on the purpose and goals of digital strategies (e.g. customers may benefit from digitizing channels while entities in the supply chain may benefit from digitizing processes). Therefore, we propose the following as a more representative dimension of digital strategies: ***Digital strategy represents the economic and*** ***non-economic contribution to stakeholders in the dynamic ecosystem in which the firm operates where stakeholders may be co-creators of value***.

#### 4.2.6  Strategy Differentiates Managerial Tasks at the Different Hierarchical Levels

Hax (1990, p. 36) purports that "Strategy is a logical system for differentiating managerial tasks at corporate, business, and functional levels". While initial findings indicate that this dimension may be applicable in the digital context, further research is required to understand its manifestation and nuances in the context of digital strategies. Four models for deploying digital strategies were identified by Participant A, with progressively increasing governance requirements: 1) Business units formulate digital strategies with "*technology delivering it*"; 2) a "*technology shared service that serves all business units*", which will require some "*coordination of platform decisions and demand management*"; 3) a Digital Centre of Excellence and technology "*coordinate strategy*" to achieve "*digital consistency across business units*"; and, 4) a digital transformation that is all-encompassing that provides direction to business units on all necessary elements to carry out the digital transformation. It appears that the model adopted and the extent of governance required depends "*on the extent to which digital is important*" to an organization, the way in which it is structured (Participant A), the purpose of its digital strategy, and who has the necessary competencies and are best placed to deliver the required outcomes. Digital strategies may evolve under the Chief Operations Officer if cost pressures exist and the organization's attempting to digitize its operations, Chief Marketing Officer if it's driven by marketing and the entity's seeking to enhance the customer experience or Chief Financial Officer if the entity's seeking to transform business and revenue models (Participant B). An evolution of new roles such as a Chief Digital Officer is evident, perhaps resulting from the need to "*bring in somebody with the capability*" (Participant B). According to Bharadwaj et al. (2013, p. 473), digital business strategy "transcends traditional functional areas" and will evolve into an entity's business strategy as digitization proliferates. However, to associate a digital strategy solely with a business strategy may not be true in all contexts as Theme 8 in the previous section indicates that digital strategies could be an enabler and a crucial element in an overarching corporate strategy.

### 4.3  Additional Dimensions of Strategy

In the previous section, we discussed the concept of digital strategy development and execution being an iterative and experimental process (Theme 6) and the need for digital strategy development and implementation to be agile and responsive (Theme 7). Therefore, we propose two new dimensions to extend the definition of strategy proposed by Hax (1990) to reflect the nature of digital strategies. Further research may be undertaken to explore the applicability of these findings to conventional strategies.

#### 4.3.1  Digital Strategy Development and Execution as Agile and Responsive

According to Pagani (2013, p. 620), "in order to succeed over the long haul, firms within value networks have to periodically reorient themselves by adopting new strategies and structures that are necessary to accommodate changing environmental conditions." Agility may be achieved through greater speed (e.g. of product releases, of decision-making facilitated by big data analytics, of "supply chain orchestration" and/or of "network formation and adaptation"), and the ability to dynamically scale digital capabilities e.g. through the adoption of a cloud computing model and the development of strategic partnerships (Bharadwaj et al. 2013, pp. 475-477).

#### 4.3.2  Digital Strategy Development as an Iterative and Experimental Process

This is a new dimension of digital strategy that was identified from the interview data that does not form a part of Hax's definition of strategy. While a study conducted by the MIT Center for Digital Business and Capgemini Consulting (2011) indicates that digital strategy may be an iterative and experimental process, the research appears to be oriented towards practitioners with little evidence of scientific and methodological rigour. We are not aware of any previous academic studies that explicitly link the concept of 'digital strategy development' with the notion of 'iterative' and 'experimental'.

## 5  CONCLUSION, LIMITATIONS AND FUTURE WORK

The main purpose of this research-in-progress paper was to explore the various conceptions of digital strategy and provide insight into how digital strategies differ from conventional strategies. We identified various themes from the interview data, grouping them under one of three theoretical constructs. Additional interviews will be conducted to gain further insight into digital strategies and provide a greater understanding of the nuances in the identified themes and the social context that influenced conceptions. Future work will also focus on clearly establishing the difference between an IT strategy and a digital strategy.

We applied the understanding of digital strategies from the interview data and existing literature to the six dimensions of strategy proposed by Hax (1990) to explore the applicability of generic strategy constructs to digital strategies. We asserted that digital strategy **establishes an organization's long-term goals and objectives for digitization**. We found that digital strategy defines **the dynamic ecosystem in which an entity operates or intends to operate** as opposed to classification of an entity's competitive domain based on traditional industry silos. Digital strategy may be a response to external opportunities and threats and to internal strengths and weaknesses **to achieve competitive advantage through successive ephemeral advantages**. We also proposed an alternative view of this dimension based on the interview data: Digital strategy may be perceived as a response to external opportunities and threats and internal strengths and weaknesses **to sustain an entity's competitive position**. Based on the findings from the interview data, we proposed two new dimensions to extend Hax's definition to the digital context: Digital strategy development and execution as **agile and responsive**; and digital strategy development and execution as **an iterative and experimental process**. Future research could explore the applicability of these new dimensions of digital strategy to strategy development in other contexts. The governance implications relating to digital strategies being an enabler for potentially all levels of strategy in organisations may also be explored in subsequent studies.

This appears to be the first detailed study that seeks to explore qualitatively, the different ways in which digital strategies may be conceived, thereby providing a robust foundation for future academic research in developing prescriptive frameworks for digital strategy development and implementation. This paper provides a greater understanding of digital strategies and explores the way in it differs from conventional strategies, which will highly benefit practitioners in implementing digital strategies. The comparison of digital strategy and conventional strategy highlights that a different approach may be required to develop and implement strategy in an increasingly digital world. The knowledge that digital strategy development is an iterative and experimental process that necessitates agility and responsiveness to environmental changes, such as in the technological landscape and consumer expectations, provides guidance and assurance to consultants in their approach to digital strategy development. A deeper understanding of digital strategy may be obtained from these research findings and verified through additional interviews. We expect this will provide greater insight into permutations within each of the identified dimensions and facilitate a unified understanding of digital strategies to support further academic research and provide guidance to practitioners on how digital strategies may be understood and approached.

A limitation of this study is that the executives interviewed to date primarily have a business or IT background. Thus, the findings may not reflect other perspectives such as that of digital marketing executives, for instance. Additional interviews are being organised to address this gap. Furthermore, this study is focused on perceptions of digital strategy held by entities in traditional, medium to large organisations.

## Copyright